\documentclass{mn2e}
\usepackage{color, soul}
\usepackage{graphicx}
\usepackage{amsbsy}
\usepackage{amsmath}
\usepackage{epsf}
\usepackage{paralist}
\usepackage{epstopdf}
\usepackage{amssymb}
\usepackage{multirow}
\usepackage{lscape}
\usepackage{titlesec}
\usepackage[flushleft]{threeparttable}
\usepackage[symbol*]{footmisc}
\newcommand{\apj}{ApJ}
\newcommand{\apjl}{ApJL}
\newcommand{\mnras}{MNRAS}
\newcommand{\apjs}{ApJS}

\newcommand{\aap}{A\&A}

\newcommand{\araa}{ARA\&A}

\newcommand{\prd}{Phys. Rev. D}

\def\be{\begin{equation}}
\def\ee{\end{equation}}
\def\bea{\begin{eqnarray}}
\def\eea{\end{eqnarray}}

\def\sp{\;\;\;\;\;\;}
\def\spa{\;\;}

\begin{document}

%===========================================================================
\title[EGRB from star-forming galaxies: do empirical scalings suffice?]
{Extragalactic Gamma-ray Background from Star-forming Galaxies: Will Empirical Scalings Suffice?}
%===========================================================================

\author[Komis, Pavlidou \& Zezas]
  {I.~Komis$^{1}$, V.~Pavlidou$^{1,2}$, A.~Zezas$^{1,2}$ \\
    $^1$Department of Physics and Institute for Theoretical and Computational Physics, University of Crete, PO Box 2208, 71003 Heraklion, Greece\\
    $^2$IESL, Foundation for Research and Technology-Hellas, PO Box 1527, 71110 Heraklion, Crete, Greece}

\maketitle

\begin{abstract}
Despite the influx of unprecedented-quality data from the Fermi Gamma-Ray Space Telescope that have been collected over nine years of operation, the contribution of normal star-forming galaxies to the extragalactic gamma-ray background (EGRB) remains poorly constrained. Different estimates are discrepant both their underlying physical assumptions and their results. With several detections and many upper limits for the gamma-ray fluxes of nearby star-forming galaxies now available, estimates that rely on empirical scalings between gamma-ray and longer-wavelength luminosities have become possible and increasingly popular. In this paper we examine factors that can bias such estimates, including: a) possible sources of nontrivial redshift dependence; b) dependence on the choice of star-formation tracer; c) uncertainties in the slope and normalisation of empirical scalings. We find that such biases can be significant, pointing towards the need for more sophisticated models for the star-forming galaxy contribution to the gamma-ray background, implementing more, and more confident, physics in their buildup. Finally, 
we show that there are large regions of acceptable parameter space in observational inputs that significantly overproduce the gamma-ray background, implying that the observed level of the background can yield significant constraints on models of the average cosmic gamma-ray emissivity associated with star formation. 
\end{abstract}

\begin{keywords}
gamma-rays: diffuse background -- gamma-rays: galaxies -- methods: analytical -- galaxies: luminosity function, mass function -- 
\end{keywords}

\section{Introduction} 
The gamma-ray sky consists of resolved point sources, Galactic diffuse emission, i.e. photons produced in interactions between cosmic rays and the interstellar medium and interstellar radiation field, and an isotropic, presumably extragalactic diffuse emission, the extragalactic gamma-ray background (EGRB, Clark et al. 1968; Fichtel et al. 1978; Sreekumar et al. 1998; Strong et al. 2004; Abdo et al. 2010a; Ackermann et al. 2015). The EGRB is a superposition of individual unresolved point sources and truly diffuse emission, and it encodes valuable information about high energy processes in the universe.  The dominant components of the EGRB are likely active galaxies (blazars), which are the brightest extragalactic sources and the most numerous population of resolved gamma-ray sources, and normal star-forming galaxies (Pavlidou \& Fields 2002; Fields, Pavlidou \& Prodanovi\'{c} 2010; Abdo et al. 2010b; Stecker \& Venters 2011). These two classes of gamma-ray sources are guaranteed to contribute to the EGRB. 

The relatively large number of resolved gamma-ray blazars has made possible the construction of empirical gamma-ray luminosity functions, which, when extrapolated to fainter fluxes, place the contribution of unresolved blazars around $\sim 50\%$ of the Fermi LAT measurement of the EGRB (Ajello et al. 2015).  Star-forming galaxies on the other hand are individually faint in gamma rays, and only a handful of sources have been individually resolved (Ackermann et al. 2012b; Tang et al. 2014; Peng et al. 2016). As a result, the construction of empirical gamma-ray luminosity functions remains infeasible, and indirect ways of estimating their contribution to the EGRB have to be employed. 

In normal star-forming galaxies the dominant emission mechanism of gamma-rays is through hadronic interactions between cosmic rays (CR) and interstellar gas (IG) producing neutral pions, i.e., $p_{\mbox{\scriptsize{CR}}} + p_{\mbox{\scriptsize{IG}}} \to pp{\pi ^0}$. Gamma-rays are produced through pion decay $\pi ^0 \to \gamma \gamma$ (Stecker 1971). A critical assumption in the case of normal star-forming galaxies is that the level of cosmic ray flux is set by the balance between cosmic-ray acceleration (likely in supernova remnants) and diffusive escape of cosmic rays to the intergalactic medium (e.g., Pavlidou \& Fields 2001). In order for this assumption to hold, cosmic-ray losses due to escape have to dominate over losses due to pion production. In this case, the flux of gamma rays depends on: a) the flux of projectiles (cosmic rays) and b) the number of targets (IG particles). Since both these quantities are related to the star formation rate (SFR) of a galaxy, the gamma-ray emissivity of a galaxy is expected to depend on its star-formation properties. 

In contrast to normal star-forming galaxies, starburst galaxies tend to have much higher gas fractions, and losses due to pion production may dominate. In this case, the flux of cosmic rays is set by the balance between acceleration and pion production, and the flux of gamma rays is simply proportional to the cosmic ray acceleration rate. The gamma-ray emissivity of such a {\em calorimetric} galaxy will still depend on its star formation rate, but following a different scaling than that of normal galaxies (e.g., Thompson et al. 2007, Wang \& Fields 2016). For this reason, normal and starburst galaxies are treated as distinct classes with respect to their contribution to the EGRB.

Before the launch of Fermi, the only star-forming galaxies with confirmed gamma-ray emission were the Milky Way and the Large Magellanic Cloud. Of necessity, then, early studies of the star-forming galaxy contribution to the EGRB relied on theoretical arguments (e.g., Pavlidou \& Fields 2002; Fields et al. 2010; Makiya et al. 2011; Stecker \& Venters 2011). However, Fermi has now made possible the detection of several nearby star-forming galaxies, and provided upper limits for the gamma-ray flux of many more (Ackermann et al. 2012b). 
As a result, it is now possible to obtain empirical scalings between gamma-ray luminosity and luminosities at lower, star-formation-tracing frequencies, and use such scalings together with lower-frequency luminosity functions to obtain estimates of the EGRB contribution from star-forming galaxies (e.g., Stecker \& Venters 2011; Ackermann et al. 2012b; Linden 2017). 

However, there is still no consensus among these different studies on the fraction of the EGRB that can be attributed to star-forming galaxies. The difference is not just arithmetical: rather, there is tension between the theoretically expected and the empirically observed slope of the correlation between SFR and gamma-ray luminosity ($L_{\gamma}$). While Fields et al. (2010), using the Kennicutt-Schmidt (KS) relation, find a slope of $\sim 1.7$, Ackermann et al. (2012b) and Linden (2017) report a best-guess slope closer to $\sim 1.2$. 
Given that star-forming galaxies are very faint in gamma rays, it is not expected that enough members of this class will be individually resolved in the foreseeable future to construct empirical luminosity functions. It is therefore important to understand the origin of any such tensions; to investigate whether empirical scalings and longer-wavelength luminosity functions can act as a reliable substitute of empirical luminosity functions; and identify any potential biases in the construction of models for the star-formation-related EGRB components from empirical scalings, as well as ways to address and remedy such biases. This is the scope of this paper. 

Here, we examine two factors as potential sources of bias in EGRB estimates of star-forming galaxies based on empirical scalings between gamma-ray luminosity and longer-wavelength luminosities. First, possible sources of nontrivial redshift dependence, such as the dependence of the supernova progenitor mass cutoff on metallicity, or a redshift-dependent performance of different star formation tracers. And, second, uncertainties in the slope and normalisation of empirical scalings, arising either due to the extremely-low-number statistics involved, or due to the evolution with cosmic time of the physical processes setting these scalings. 

This paper is organised as follows. In \S \ref{model} we describe the general formalism for a luminosity-function--based calculation of the contribution of a certain class to the EGRB. In \S \ref{lf} we discuss how the different factors discussed above enter the construction of a gamma-ray luminosity function. Our results are presented in \S \ref{results}, and they are discussed in the context of recent work in gamma-ray astrophysics and the cosmic history and star formation in \S \ref{Disc}, where a roadmap towards future improvements in the modeling of gamma-ray emission for star-forming galaxies is also suggested. 
% in Section 2, we discuss the different models we used to derive the contribution of the normal star-forming galaxies to the EGRB as well as the possible sources of discrepancy between the theoretical and the empirical relation. In Section 3, we present our results of the contribution of normal SFGs and the bounds to the potential contribution of star forming galaxies to the EGRB. Finally, in Section 4, we present our conclusions.  

\section{Formalism} \label{model}
The diffuse gamma-ray photon intensity produced by primary emission from unresolved star-forming galaxies with luminosity function $\Phi ({L_\gamma },z) = dn/dL_\gamma$, where $n$ is the comoving number density, is given by the the line-of-sight integral over the gamma-ray luminosity function:
\begin{equation}\label{main}
I_E(E) = \int\limits_0^{z_{\max }}\!\!dz \!\!\!\!\int\limits_{L_{\gamma ,\min }}^{L_{\gamma ,\max }} \!\!\!\!\!dL_\gamma 
\Phi ({L_\gamma },z)\frac{dV}{dzd\Omega } F_1\left[L_\gamma, E(1 + z), z\right]
\end{equation}
where $E$ is the observed photon energy, $F_1[ L_\gamma, E(1 + z), z]$ 
is the differential photon flux of an individual galaxy of redshift $z$ and gamma-ray luminosity $L_{\gamma}$, and
$dV/dzd\Omega$ is the comoving volume element per unit redshift and unit solid angle. In Eq.~(\ref{main}) we have assumed that all galaxies have identical photon spectral shapes, and that we only treat energies where absorption of gamma rays by pair production on the intergalactic background light (see, e.g., Venters 2010) is negligible. In this work, we will express the gamma-ray luminosity function in terms of the differential photon luminosity at a rest-frame energy of 0.6 GeV since most photons come from redshifts around $z\sim1$:  $L_\gamma \equiv L_{\gamma,E}(0.6 {\rm \, GeV})$. We use this same quantity to normalize the photon spectrum,
$L_{\gamma,E} (E) = L_\gamma dN/dE$, where $dN/dE|_{\rm 0.6\, GeV}=1$. The photon flux from a single galaxy is then given by 
\begin{equation}
F_1[ L_\gamma, E(1 + z), z] = \frac{L_\gamma}{4\pi \xi^2}\frac{dN}{dE} \left[E(1+z)\right]\,
\end{equation}
where $\xi$ is the coordinate distance that enters the Robertson-Walker metric, 
\begin{equation}
ds^2 = -c^2dt^2 +a^2(t)\left[d\xi^2/(1-k\xi^2) + \xi^2d\Omega^2\right]
\end{equation}
%In general, the comoving volume element per unit solid angle and per unit redshift is 
%\begin{equation}
%\left|\frac{dV}{d\Omega} \right|= \frac{c}{H_0}\frac{\xi^2}{\sqrt{\Omega_m (1+z)^3 + \Omega_\Lambda + %(1-\Omega_m-\Omega_\Lambda)(1+z)^2}}
%\end{equation}
so for a flat Universe $(k=0, \Omega_m+\Omega_\Lambda = 1)$ 
 the comoving volume element per unit solid angle and per unit redshift is 
%\begin{equation}
$|dV/d\Omega dz|= (c\xi^2)/(H_0\sqrt{\Omega_m (1+z)^3 + \Omega_\Lambda})$
%\end{equation}
so Eq.~(\ref{main}) simplifies to
\begin{equation}
I_E(E) = \frac{c}{4\pi H_0}\!\!\!\int\limits_0^{z_{\max }}\!\!dz \!\!\!\!\int\limits_{L_{\gamma ,\min }}^{L_{\gamma ,\max }} \!\!\!\!\!\, \frac{dL_\gamma
 \Phi ({L_\gamma },z)  L_\gamma}{\sqrt{\Omega_m(1+z)^2+\Omega_\Lambda}} \frac{dN}{dE} \left[E(1+z)\right].
\end{equation}
We use a standard $\Lambda$CDM cosmology with ${\Omega _M} = 0.3, \spa {\Omega _\Lambda } = 0.7, \spa {H_0} = 73 {\rm \,\, km\,\,sec^{ - 1} \,Mpc^{ - 1}}$.

\section{The Gamma-ray Luminosity Function}\label{lf}
Since $\Phi(L_\gamma,z)$ cannot be determined directly from Fermi data (because resolved star-forming galaxies are too few), it is common practice to rescale an infrared luminosity function assuming some relation between gamma-ray luminosity $L_{\gamma}$ and infrared luminosity $L_{\rm IR}$. A very significant fraction of the uncertainty in the calculation of the contribution of normal star-forming galaxies to the EGRB enters through the assumed relation between $L_\gamma$ and $L_{\rm IR}$. There are various approaches to deriving such a relation, which lead to different results.
 
\subsection{Empirical scaling between $L_\gamma$ and $L_{\rm IR}$}
Ackermann et al. (2012b) optimised simple power law scalings between gamma-ray luminosity (energy luminosity integrated between 100 MeV and 100 GeV, $L_{0.1 - 100 \rm GeV}$) and total infrared luminosity, $L_{8-1000 \rm \mu m}$:
\begin{equation}\label{empirical}
\log \left( \frac{L_{0.1 - 100 \rm \, GeV }}{\rm erg \,\, s^{ - 1}} \right) = \alpha \log \left( {\frac{L_{8 - 1000\mu\rm m}}{{{{10}^{10}}{L_ \odot }}}} \right) + \tilde \beta\,.
\end{equation}
In the equation above, the energy luminosity integrated between 0.1 and 100 GeV that enters the scaling is related to the differential photon luminosity at 0.6 GeV, $L_\gamma$, that enters Eq.~(\ref{main}) through $L_{0.1 - 100 \rm \, GeV} = \int_{E=0.1 \rm \, GeV}^{100 \rm \, GeV} E L_\gamma (dN/dE)dE$. 

Ackermann et al. (2012b), depending on whether galaxies with AGN were included in (excluded from) their analysis, found a slope $\alpha \sim 1.17$ ($\sim 1.1$), using the Expectation-Maximization (EM) algorithm [e.g., Dellaert (2002) and references therein], which is similar to the well known least-square fitting. A more sophisticated statistical analysis by Linden (2017) taking explicitly into account the possible spread in the optimised scaling, also resulted in a consistent result ($\sim 1.18$) for the slope. These analyses take into account a significant number of star-forming galaxies for which only upper limits, rather than statistically significant measurements, are available for their gamma-ray flux. However, most of these upper limits are weak, and we have confirmed that as a result the best-fit value for the scaling slope is naturally dominated by the effect of the (few) resolved galaxies. Performing a power law fit using only resolved galaxies we found same slope as the analysis including the upper limits. For this reason, in our own tests for sources of biases entering through an adopted empirical scaling, we limit ourselves to least-square fitting of power laws using the resolved sources only. 

In using Eq.~(\ref{empirical}) and an infrared luminosity function to obtain the gamma-ray luminosity function, we adopt the implicit assumption that the scaling itself does not show a non-trivial redshift evolution. A few ways such a redshift dependence could enter, include, for example, the following.
\begin{itemize}
\item[a)] In escape-dominated galaxies, where the gamma-ray flux depends both on the  gas mass and the SFR, the gas fraction of a galaxy of given mass increases with redshift, compounding the effect of increased star formation. 
\item[b)] The performance of infrared flux as a star-formation tracer is known to be  redshift-dependent, so that infrared luminosity functions may not adequately represent the cosmic star formation history (which in turn sets the history of cosmic ray acceleration and thus star-formation-related gamma-ray emission). 
\item[c)] The supernova mass cutoff is metallicity-dependent, so that at high redshift / lower metallicity environments, a different fraction of stars act as supernova progenitors that will subsequently produce supernova remnants that will participate in cosmic-ray acceleration.
\item[d)] The fraction of star formation that takes place in environments so gas-rich that gamma-ray production is calorimetric may increase with increasing redshift. This may impart a redshift-dependent change is the scaling slope between gamma-ray and infrared luminosities. 
\end{itemize}
In this paper we quantitatively assess effects (a)-(c) above, while we qualitatively discuss the potential impact of (d) and ways it can be addressed in the future. 

\subsection{Analytically derived scaling between $L_\gamma$ and $L_{\rm IR}$}
An alternative approach to relating $L_\gamma$ and $L_{\rm IR}$ is to assume that $L_{\rm IR}$ is proportional to the SFR, and then use our understanding of the physics of gamma-ray production in star-forming galaxies to relate the SFR to $L_\gamma$. This is the approach used by Fields et al. (2010).  

For galaxies with escape-dominated cosmic-ray losses,  the gamma-ray production rate per interstellar H-atom, $\Gamma_{\pi^0 \rightarrow \gamma \gamma}$, is proportional to the galaxy's volume averaged cosmic-ray proton flux, $\Phi_{\rm cr}$ (e.g., Stecker 1977; Pohl 1994; Pavlidou \& Fields 2001; Persic \& Rephaeli 2010), so 
\be
L_{\gamma} \propto M_{\mbox{\scriptsize{gas}}} \Phi _{\mbox{\scriptsize{cr}}}
\ee
where $M_{\mbox{\scriptsize{gas}}}$ is the gas mass of a galaxy and $\Phi _{\mbox{\scriptsize{cr}}}$ its cosmic-ray flux.
If in addition we assume that the initial mass function (IMF), the supernova progenitor mass cutoff,  and the confinement efficiency are comparable in all galaxies, while all supernova remnants accelerate, on average, the same number of cosmic rays\footnote{In reality, all of these quantities may show variations, but as long as they are not dependent on star formation rate or gas content of a galaxy, they will simply contribute to the scatter of the relation, or, if they happen to evolve with cosmic time, introduce a redshift dependence. }, 
\be
\Phi_{\rm cr} \propto R_{\rm SN} \propto \psi
\ee
where $R_{\rm SN}$ is the supernova rate (SNR) and $\psi$ is the SFR. Thus,
\be
L_{\gamma} \propto M_{\mbox{\scriptsize{gas}}} \psi
\ee

$M_{\rm gas}$ can also be related to $\psi$. For example, the Kennicutt-Schmidt scaling (Schmidt 1959; Kennicutt 1998) relates the SFR and gas surface densities,
\be 
\Sigma_{\mbox{\scriptsize{SFR}}}   \propto \Sigma_{\mbox{\scriptsize{gas}}}^x
\ee
which yields (Fields et al. 2010) 
\be
{M_{\mbox{\scriptsize{gas}}} (\psi, z)} \propto {(1 + z)^{ - \beta }} \psi^\omega
\ee
where $\beta  = 2(1 - 1/x)$ and $\omega  = 1/x$. The $(1 + z)^{- \beta}$ factor enters through the conversion of surface densities of gas and SFR in the KS law to volume densities. The gamma-ray luminosity of a galaxy then will be,
\be
{L_\gamma }\left( {\psi ,z} \right) \propto (1 + z)^{ - \beta } \psi^{\omega  + 1}\,.
\ee
Note that this equation is valid only for normal escape-dominated galaxies. In starburst galaxies, which have very high cosmic-ray intensities and gas within small volumes, inelastic collisions compete with and sometimes even dominate over escape to regulate cosmic-rays losses (Paglione et al. 1996; Torres et al. 2004; Stecker 2007; Thompson et al. 2007; Lacki et al. 2011; Persic \& Rephaeli 2010; Wang \& Fields 2016).  It is important then to keep in mind that unless we explicitly exclude starburst galaxies from our calculation, we will end up overestimating the gamma-ray signal. 

The total infrared luminosity is a well-established tracer of the SFR for late type galaxies (Kennicutt 1998a). The conversion proposed by Kennicutt (1998b) is the following:
\be\label{scaling}
\frac{\psi }{{1{M_ \odot }{\mbox{{yr}}^{ - 1}}}} = \epsilon 1.7 \times {10^{ - 10}}\frac{{{L_{8 - 1000\mu m}}}}{{{L_ \odot }}}
\ee
This conversion assumes that thermal emission of interstellar dust approximates a calorimetric measure of radiation produced by young, i.e. $10-100$ Myr, stellar populations. The factor $\epsilon$ depends on the assumed initial mass function (IMF). Throughout this work we use Salpeter IMF (Salpeter 1955) and we consider it unchanging through space and time. 
Hence, the scaling relation between gamma-ray luminosity and total infrared luminosity (TIR) is
\be\label{final}
{L_ \gamma} (L_{8 - 1000\mu m}, z) \propto {(1 + z)^{ - \beta }}{\left(\frac{{{L_{8 - 1000\mu m}}}}{{{L_ \odot }}} \right)^{\omega  + 1}}
\ee
In order then to calculate $\Phi (L_{\gamma}, z)$, which enters Eq. (\ref{main}), we need to (1) adopt a luminosity function, (2) determine the slopes $\beta$ and $\omega$ (either from KS scaling or empirically) and (3) determine the normalisation of the scaling in Eq.~(\ref{final}). The latter can be derived using, for example, data from the Milky Way, where the local cosmic-ray flux, gas mass, and gamma-ray emission are well measured, or observations of nearby galaxies resolved in gamma rays. 

In the model of Ackermann et al. (2012b) the gamma-ray data used in order to derive the scaling relation between the gamma-ray luminosity and TIR luminosity of a galaxy include, in addition to upper limits, eight resolved galaxies. However, two of them (NGC 4945 and NGC 1068) are galaxies with active galactic nuclei (AGN) so, we are not going to consider them in this work. Our sample of galaxies consists of four normal star-forming galaxies (SMC, LMC, MW, M31) and four starburst (NGC 253, M82, NGC 2146, and Arp 220); effects of including the starburst galaxies in deriving the slope and normalisation of the scaling will be discussed in detail below. Detections of NGC2146 (Tang et al. 2014) and Arp 220 (Peng et al. 2016) were not available at the time of the Ackermann et al. (2012b) analysis. 

\subsubsection{Infrared Luminosity Function}

We begin by considering the adopted luminosity function. Fields et al. (2010) use an H$\alpha$ luminosity function while Ackermann et al. (2012b) use the luminosity function of Rodighiero et al. (2010):

\be\label{rod}
\begin{split}
&\Phi (L){\mbox{d}}{\log _{10}}(L) =  \\
&{\Phi ^ * }{\left( {\frac{L}{{{L^ * }}}} \right)^{1 - \gamma }}\exp \left[ { - \frac{1}{{2{\sigma ^2}}}\log _{10}^2\left( {1 + \frac{L}{{{L^ * }}}} \right)} \right]{\mbox{d}}{\log _{10}}(L)
\end{split}
\ee
where, the parameter $\gamma$ (defined as $\alpha$ in the paper of Rodighiero et al. (2010)) sets the slope at the faint end, $L^*$ is the characteristic luminosity and $\Phi ^*$ is the normalisation factor.

The choice of a luminosity function in some star-formation-tracing frequency is the first point of divergence between the  different models: each choice sets an (explicit) redshift dependence of the gamma-ray emissivity, even if all other elements of a model remain the same. However, this effect can be controlled and corrected in a straight-forward way, described below. 

\begin{figure}
\includegraphics[width=1.0\columnwidth, clip]{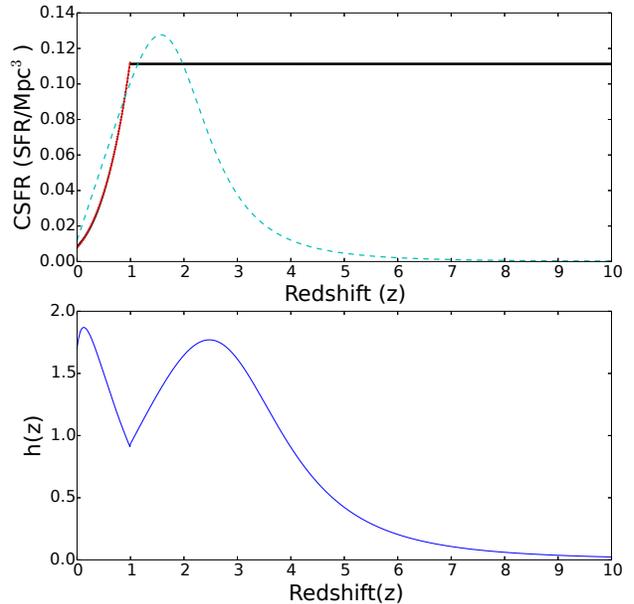}
\caption{\textit{Upper:} \textit{cyan dashed} line: ${{\dot \rho }_*}$ Hopkins \& Beacom (2006) cosmic star-formation history. \textit{Black} line: cosmic star-formation history derived from the luminosity function of Rodighiero et al. (2010), assuming no decline beyond $z=1$. \textit{Red} line: our fit to the cosmic star formation history derived from the Rodighiero et al. (2010) luminosity function for $z\le 1$.
\textit{Bottom:} The correction factor $h(z)$ we have introduced to match the star-formation history of Hopkins \& Beacom (2006). \label{hz}}
\end{figure}

A luminosity function in some star-formation tracer yields a history of cosmic star formation rate density, once we convert luminosity to star formation rate and integrate over all luminosities. Conversely, if we decide on a ``preferred'' cosmic star formation history, we can adjust any luminosity function using an overall redshift-dependent multiplicative factor so that, once integrated, it reproduces the desired cosmic star formation history. This approach has the advantage that it minimizes the sensitivity to biases relevant to each tracer. Computing a star-formation history using multiple tracers effectively averages over several different indicators thus generating a more robust result.

This is the approach we demonstrate and use in this work. As our ``preferred'' choice we use the star-formation history ${{\dot \rho }_*}$ as surmised by Hopkins \& Beacom (2006), considering all available SFR tracers. The luminosity function of Rodighiero et al. (2010) (which we choose to use for the remainder of this work) is not in agreement with this result (for example, Rodighiero et al. 2010 find no significant redshift dependence for $z>1$, while Hopkins \& Beacom 2006 report a strong ${{\dot \rho }_*}$ peak at $z=1.5$ and a sharp decline at higher $z$). To bring the two in agreement, we introduce a redshift-dependent dimensionless normalisation correction factor in the scaling of Eq.~(\ref{scaling}):
\be
\frac{\psi }{{1{M_ \odot }{\mbox{{yr}}^{ - 1}}}} = \epsilon 1.7 \times {10^{ - 10}} h(z) \frac{{{L_{8 - 1000\mu m}}}}{{{L_ \odot }}}
\ee
such that, $\int {\psi \Phi (\psi ,z){\mbox{d}}{{\log }_{10}}\psi  = {{\dot \rho }_*}}$, where ${{\dot \rho }_*}$ is the Hopkins \& Beacom (2006) cosmic star-formation history, and $\Phi (\psi, z)$ is the star-formation rate distribution function obtained by the luminosity function of Eq.~(\ref{rod}) and the scaling of Eq.~(\ref{scaling}). Fitting an analytic expression to $\int \psi \Phi (\psi ,z) {\rm d}\!\log _{10}\psi$, we obtain:
\be
h(z) = \left\{ \begin{array}{l}
0.7\frac{{0.017 + 0.13z}}{{\left[ {1 + {{\left( {\frac{z}{{3.3}}} \right)}^{5.3}}} \right]\left( {0.022\exp (1.77z) - 0.015} \right)}} \spa \; 0 < z < 1 \\
6.36\frac{{0.017 + 0.13z}}{{1 + {{\left( {\frac{z}{{3.3}}} \right)}^{5.3}}}}  \sp \sp \sp \sp \sp \sp  \spa  z > 1
\end{array} \right..
\ee
In Fig.~\ref{hz} (lower panel)  we plot $h$ as a function of $z$. Despite its complicated functional form, $0.5 < h(z) < 2$ for a large range of redshifts (all $z<4$).

This approach allows for a fair comparison between models using different luminosity functions, but also between models that are luminosity-function--based (including, e.g., the Fields et al. 2010, Stecker \& Venters 2011, and Ackermann et al. 2012b models) and models that are based on the overall cosmic star formation history (such as the Pavlidou \& Fields 2002 and the Ando \& Pavlidou 2009 models). 

The correction factor $h(z)$ is the first redshift dependence we identify (cosmic evolution of the performance of a single star formation tracer)
 that is unaccounted for when using empirical scalings alone. 

\subsubsection{Scaling slopes $\omega$ \& $\beta$}
Perhaps the most puzzling discrepancy between the theoretical approach of Fields et al. (2010) and the empirical scalings of Ackermann et al. (2012b) and Linden (2017) is the discrepancy in the scaling slope $\omega + 1$ between $L_\gamma$ and $\psi$.

From a physics perspective, for escape-dominated galaxies, if $L_{8-1000 \mu m}$ is indeed proportional to $\psi$ 
%(and is not also modulated by the gas content of a galaxy), 
the $L_{\gamma} - L_{8-1000 \mu m}$ scaling slope {\emph{should}} deviate significantly from unity to reflect the compounded effect of both star-formation ($\rightarrow$ cosmic ray accelerators $\rightarrow$ flux of projectiles) and gas ($\rightarrow$ availability of targets). On first inspection, this could be the effect of including starburst galaxies in the dataset from which the slope is calculated: starbursts are expected to be calorimetric and hence to exhibit a $L_{\gamma} \propto L_{8-1000 \mu m}$ scaling (see e.g. Wang \& Fields 2016); since Ackermann et al. (2012b) and Linden (2017) consider both normal and starburst galaxies when determining a best-guess slope, it is reasonable to expect a slope closer to 1 than when considering normal galaxies alone. However, this is not the case: when fitting only the normal star-forming galaxies detected by Fermi (Milky Way, LMC, SMC, and M31), we find a slope of 0.9, even flatter than the $\sim 1.2$ slope derived for all galaxies!

If the empirical scaling for normal galaxies alone does indeed exclude a steeper slope, then this would have important implications: it could imply, for example, that confinement of cosmic rays in normal galaxies is not only variable, but SFR-dependent (with higher SFR galaxies exhibiting poorer confinement properties); or that the IMF is SFR-dependent; or that any scatter in the $L_{8-1000 \mu m} - \mbox{SFR}$ scaling is dependent on gas content; or finally, that the primary contribution in the $\gamma - \mbox{ray}$ flux from star-forming galaxies is leptonic rather than hadronic and thus dependent on cosmic-ray flux but not on gas.

Before such scenarios are entertained, however, we need to test the extent to which the scaling slope is well-constrained. There are two important systematic effects that could bias the best-fit slope more than what statistical uncertainties allow for. 

The first systematic effect concerns the choice of star formation tracer. For example, it has been suggested that the sum of near-ultraviolet (NUV) plus $25\mu m$ luminosity of each galaxy ($\nu L_{\mbox{\scriptsize{NUV}}} + 2.26L_{25 \mu m}$) is a better star-formation tracer than the total infrared luminosity $L_{8-1000 \mu m}$, as the $25 \mu m$ luminosity term corrects for possible existence of dust (Kennicutt \& Evans 2012). Indeed, we find that for normal galaxies, the best-guess slope in the $L_{\gamma}-  \left( \nu L_{\mbox{\scriptsize{NUV}}} + 2.26L_{25 \mu m} \right)$ scaling is $1.27$, significantly steeper than the $0.9$ slope obtained when using $L_{8-1000 \mu m}$. However, because the NUV luminosity of the Milky Way cannot be measured, this is not a fair comparison, as we have to omit the Milky Way from our fit. 

To remedy this, and improve the statistics of the fit (as, if we exclude the Milky Way, there are only three normal galaxies detected by Fermi), we use the following sample to compare best-fit slopes: (SMC, LMC, M31, NGC253, M82). These are all galaxies resolved by Fermi and used by Ackermann et al. (2012b), regardless of their starburst status, but excluding the Milky Way as well as the ones that appear to host an active galactic nucleus. We compare best-fit slopes of the scaling between  $L_{\gamma}$ and three star-formation tracers: $L_{8-1000 \mu m}$, $\left( \nu L_{\mbox{\scriptsize{NUV}}} + 2.26L_{25 \mu m} \right)$, and $\left( \nu L_{\mbox{\scriptsize{NUV}}} +0.27L_{8-1000 \mu m} \right)$ (see Hao et al. 2011). The best-fit slopes are $1.1$, $1.4$, and $1.6$ respectively, while the statistical error on the fitted slope is in all cases $\sim 0.1$. The scatter between these slope values is significantly larger than this statistical error on the fitted slope, and it is therefore clear that the choice of star-formation tracer does affect the empirically determined scaling slope.

The second effect involves the fact that the $L_{\gamma} -\psi$ scaling itself will have finite scatter. We are sampling this scaling with very few detected objects, while the weak upper limits in other nearby galaxies do not add significant constraints either on the slope or the scatter of the relation. In this way, we suffer from a form of ``cosmic variance'': if some of the very few galaxies we did happen to be able to ``draw'' in the nearby universe lie in the ``outskirts'' of the scaling, the fitted slope may appear significantly steeper or shallower than the true one. 

To test this effect, we performed Monte Carlo calculations to examine how the slope derived from a power-law fit is affected if we only fit a handful of sources drawn from a scaling with significant scatter. We assume that the scaling now takes the form [compare with Eq.~(\ref{empirical})]
\begin{equation}\label{scatter}
\log \left( \frac{L_{0.1 - 100 \rm \, GeV }}{\rm erg \,\, s^{ - 1}} \right) = a \log \left( {\frac{L_{8 - 1000\mu\rm m}}{{{{10}^{10}}{L_ \odot }}}} \right) + b +c\,.
\end{equation}
where $a$ is the true slope (which we assume to be $a=1.25$), $b$ is the normalisation of the scaling, and $c$ is a random number, which we assume to be Gaussianly distributed around zero. The standard deviation of the Gaussian quantifies the amount of scatter in the scaling.  We draw eight points from this scaling (equal to the number of Fermi-resolved star-forming galaxies), and we fit a power law using least-square fitting. We repeat the process $10^4$ times. 

\begin{figure}
\includegraphics[width=1.0\columnwidth, clip]{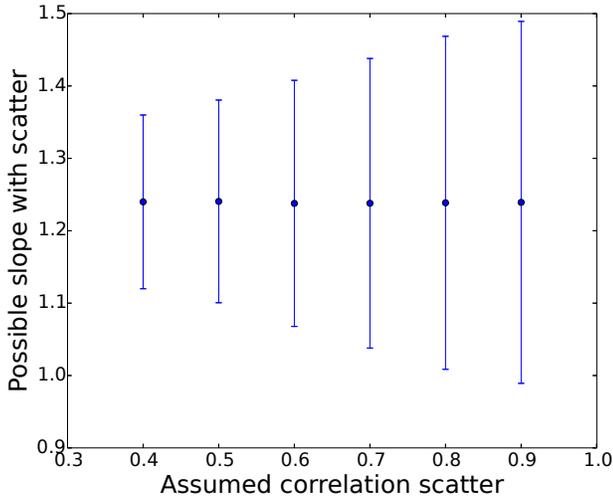}
\caption{Mean (points) and standard deviation (error bars) of fitted slopes assuming a ``true'' value of $\alpha =1.25$, and a lognormal correlation scatter (see text for details) shown in the horizontal axis. Eight datapoints used for each fit. \label{mc}}
\end{figure}

Figure \ref{mc} shows the mean and standard deviation of fitted slopes as a function of assumed scatter (standard deviation in the distribution of $c$).  
If the $L_{\gamma} - L_{\rm 8-1000 \mu m}$ scaling has a scatter of $\sim 0.4$ (as estimated, e.g., by Linden 2017), then the fitted slope carries a ``cosmic variance'' uncertainty of $\sim 0.2$, in addition to any statistical uncertainty. For scatter closer to 1 dex, the $1\sigma$ spread of slopes would be $\sim 0.5$.
Due to the two effects discussed above, at this stage a difference between $\omega + 1 = 1.7$ ($L_\gamma$ dependent on $\psi M_{\mbox{\scriptsize{gas}}}$  (Fields et al. 2010) and $\omega + 1 = 1$ ($L_\gamma$ dependent on $\psi$ alone) cannot be confidently claimed. The scaling slope between $\psi$ and $L_\gamma$ remains poorly constrained.

An additional constraint to the scaling slope may arise from the level of the EGRB itself. If an assumed value of the slope results in an overproduction of the background compared to its level established by Fermi (Ackermann et al. 2015), then such slopes would have to be excluded. To test whether such constraints are possible, we will examine both best-fit values of $\omega + 1$ using the latest data on Fermi-resolved star-forming galaxies, as well as other possible values of $\omega + 1$ between 1 (i.e., no effect of gas) and 2 (i.e., $\omega = 1/x = 1$, SFR$\propto$gas, maximum possible effect of gas). In each case, the value of $\beta$ is taken always to be consistent with that of $\omega$, i.e., $\beta = 2(1 - \omega)$. Physically, this is equivalent to changing the slope of the Kennicutt-Schmidt scaling between gas and star formation (and thus this change also propagates to $\beta$). Galaxies are still assumed to be escape-dominated. A possible convergence of the $L_\gamma - \psi$ slope towards 1 is because $\psi$ and $M_{\rm gas}$ become uncorrelated (any $\psi$ is possible for a given gas mass), rather calorimetry setting in.  

The factor $(1+z)^{-\beta}$ is the second, explicit, redshift dependence we identify (cosmic evolution of the average size of galactic disks\footnote{Note that this size evolution adds another source of uncertainty, and its validity, variance, and redshift dependence would merit their own consideration in the future.}) that is unaccounted for empirical scalings between $L_\gamma$ and $\psi$. Note that it is possible that the Kennicutt-Schmidt scaling itself evolves with redshift (see, e.g., Gnedin \& Kravtsov 2010) or with gas density (e.g. Gao \& Solomon 2004; Tassis 2007 and references therein); however we do not treat these effects here. 

\subsubsection{Normalisation of $L_{\gamma} - L_{\mbox{\scriptsize{TIR}}}$ scaling}\label{norm}
For each slope $\omega + 1$ that we will examine, we have to determine the normalisation of the scaling of Eq.~(\ref{final}). We do so by performing least-squares fitting on the sample of resolved star-forming galaxies that are relevant in each case (normal or normal + starburst), while assuming that the scaling slope is fixed at the desired value. We call the normalisation resulting in this way $L_{\gamma ,0} (\omega)$.
\begin{center}
\begin{table*}
\begin{center}
\caption{Luminosities at different wavebands of galaxies used in this work }\label{table:lums}
\begin{tabular}{c  c  c  c  c}
\hline 
\hline
Object & $L_\gamma$ & $L_{\mbox{\scriptsize{TIR}}}$ & $\nu L_{\mbox{\scriptsize{NUV}}}$ & $L_{25 \mu m}$ \\
            & $(10^{39}$erg s$^{-1})$  &  $(10^9 {L_ \odot })$ & $(10^{42}$erg s$^{-1})$ &  $(10^{22}$W Hz$^{-1})$ \\
\hline
SMC & $0.011 \pm 0.003$ [1]& $0.07 \pm 0.01$ [1]& $0.35$ [5]& $0.012$ [1] \\
LMC & $0.047 \pm 0.005$ [1]& $0.7 \pm 0.1$ [1]& $2.31$ [5]& $0.23$ [1] \\
M31 & $0.46 \pm 0.1$ [1]& $2.4 \pm 0.4$ [1]& $5.04$ [6]& $0.80$ [1] \\
Milky Way & $0.82 \pm 0.24$ [1]& $14 \pm 7$ [1]& $-$ & $7.2$ [1] \\
NGC253 & $6.0 \pm 2.0$ [1]& $21$ [1]& $1.94$ [7]& $11.6$ [1] \\
M82 & $15.0 \pm 3.0$ [1]& $46$ [1]& $1.68$ [7]& $46.0$ [1] \\
NGC2146 & $40.0 \pm 21.0$ [2]& $\sim 100$ [4]& $1.71$ [6]& $52.0$ [1] \\
Arp 220 & $1780.0 \pm 300.0$ [3]& $(1000 - 2000)$ [4]& $2.91$ [8]& $53.4$ [1] \\
\hline
\end{tabular}
\begin{center}
\begin{tablenotes}
\item References: [1] Ackermann et al.(2012b), [2] Tang et al.(2014), [3] Peng et al.(2016), [4] Gao \& Solomon(2004), [5] Kennicutt et al.(2008), [6] Gil de Paz et al.(2007), [7] Lee et al.(2011), [8] Brown et al.(2014)
\end{tablenotes}
\end{center}
\end{center}
\end{table*}
\end{center}

\subsection{Effect of Metallicity}

One of the assumptions in Fields et al. (2010) is that the ratio of cosmic-ray flux to star-formation rate is constant for all normal star-forming galaxies. In general, this will be only true on average, as the proportionality between the two depends on various properties that may be different between galaxies of the same redshift (such as the stellar IMF, the supernova progenitor mass cutoff, the acceleration efficiency in supernova remnants, or the confinement efficiency). Variations of these properties between galaxies will result in scatter in the final $L_\gamma - \psi$ scaling. 
If in addition the average value among many galaxies any of these properties evolves with redshift, then this will introduce an extra, unaccounted-for redshift dependence in the gamma-ray luminosity function. 

Here we examine the possible redshift dependence of one of these factors: the supernova progenitor mass cutoff, which depends on the metallicity, the average value of which in turn depends on redshift.  We assume that the IMF retains a Salpeter functional form (Salpeter 1955), and we explore the effect of an evolving metallicity, which can alter the minimum mass of a star that undergoes a supernovae explosion. The relation between $R_{\rm SN}$ and $\psi$ thus becomes
\be
R_{\rm SN} \propto f(Z) \psi
\ee
where $f(Z)$ encodes the effects of metallicity, $Z$. In order to specify $f(Z)$ we follow Ibeling $\&$ Heger (2013), who calculate the dependence of the low mass limit for making core-collapse supernovae on the initial stellar metallicity. Their main conclusion is that for a fixed IMF  $R_{\rm SN}$ may be $20\% - 25\%$ higher at $[Z] = -2$ than at $[Z] = 0$, where $[Z] = \log \left( {Z/{Z_ \odot }} \right)$. We are interested in the minimum mass required for a star to undergo a classical core-collapse event; following the notation of Ibeling $\&$ Heger (2013), this mass is denoted by ${M^{\mbox{\scriptsize{up}} ^{\prime} }} (Z)$. The relation they suggest for the metallicity-dependent transition mass is
\be\label{cutoff}
\frac{{{M^{\mbox{\scriptsize{up}} ^{\prime}} (Z)}}}{{{M_ \odot }}} = 
\left\{ 
\begin{gathered}
\sum\limits_{i = 0}^6 {{d_i}{{[Z]}^i}} \spa ~ : [Z] \ge - 8.3 \hfill \\
9.19 \sp \spa \spa \spa : [Z] <  - 8.3 \hfill \\ 
\end{gathered}  
\right\} \pm 0.15
\ee
where the coefficients $d_i$ (defined as $\alpha_i$ in the paper of Ibeling $\&$ Heger (2013)) are the best-fit parameters of a sixth-order polynomial they use to fit their numerical results. Following standard practice, we take the lowest-mass star to be $0.1 M_\odot$, and highest-mass star to be $120 M_\odot$. Hence, the fraction of stars undergoing a core-collapse supernova will be the integral of the IMF from $M^{\mbox{\scriptsize{up}} ^{\prime}} (Z)$ to $120 M_\odot$, divided by the integral of the IMF for the entire $0.1-120 M_\odot$ range.  Since in general $M^{\mbox{\scriptsize{up}} ^{\prime}} \ll 120 M_\odot$ and the Salpeter IMF scales as $M^{-2.35}$, $f(Z)$ will scale as 
\be
f(Z) = {\left( \frac{M^{\mbox{\scriptsize{up}} ^{\prime}} (Z)}{M^{\rm up \prime} (Z_{\rm today})} \right)}^{-1.35}
\ee
Following Kistler et al.(2013), we quantify the metallicity evolution with redshift using
\be\label{metallicity}
Z(z) =0.03\times 10^{-0.15z}
\ee
Hence, using Eqs.~(\ref{cutoff}) and (\ref{metallicity}), we obtain $f(z)$.

\subsection{Scaling including explicit redshift dependences}
Combining all the effects discussed above we finally obtain the following expression for the scaling relation between the gamma-ray luminosity and the total infrared luminosity of a galaxy:
\be\label{alltogether}
L_\gamma = L_{\gamma ,0} 
(1+z)^{- \beta} \left[ {h(z)}  \right]^{\omega +1} f(z)^{-1.35} {{\left( {\frac{{{L_{8 - 1000\mu m}}}}{{{L_ \odot }}}} \right)}^{\omega  + 1}}\,,
\ee
where $\omega$ may either be an assumed value or a best-fit value from the resolved galaxy dataset, while $ L_\gamma = L_{\gamma ,0} (\omega) $ is always obtained from the resolved galaxy dataset given a value of $\omega$, and $\beta$ is consistent with the value of $\omega$ [i.e., $\beta = 2(1-\omega)$].

\begin{figure}
\includegraphics[width=1.0\columnwidth, clip]{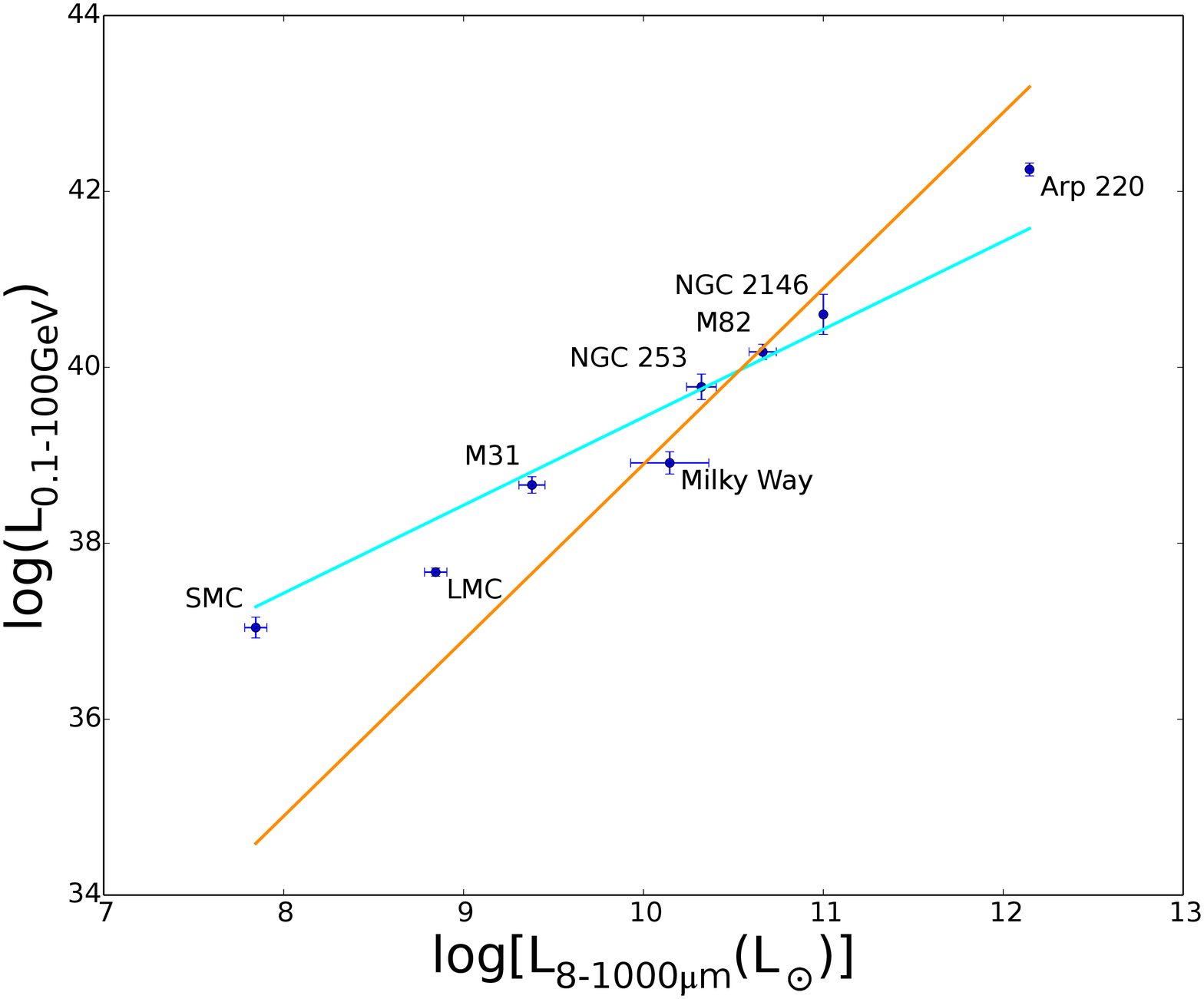}
\caption{Gamma-ray luminosity (in ergs/s) plotted against total infrared luminosity, for all galaxies considered here. The obtained scaling relations for the full sample for two choices of the scaling slope are also shown. \textit{Orange:} $\omega +1 = 2$. \textit{Cyan:} $\omega + 1 = 1$.\label{fig:scaling}}
\end{figure}

The luminosities of our sample of galaxies that we used as well as the normalization constants that we derive are shown at Table \ref{table:lums} and \ref{table:norms} respectively. In Figure \ref{fig:scaling} we have plotted two scaling relations between gamma-ray luminosity and total infrared luminosity for assumed values of $\omega + 1$ equals to 1 and 2 with cyan and orange line respectively.

\section{Results}\label{results}

\begin{table}
\begin{center}
\caption{Normalization constant for each slope value examined.}\label{table:norms}
\begin{tabular}{c  c  c  c  c}
\hline 
\hline
\multirow{3}{*}{Lines} & \multicolumn{2}{c}{Normal} & \multicolumn{2}{c}{Normal \& Starbursts} \\
 & $\omega+1$ & $L_{\gamma ,0}(\omega)$ & $\omega+1$ & $L_{\gamma ,0}(\omega)$ \\
 & &(erg s$^{-1})$& &(erg s$^{-1})$  \\
\hline
Orange & 2 & $1.86 \times 10^{20}$ & 2 & $7.89 \times 10^{18}$ \\
Cyan     & 1 & $9.88 \times 10^{28}$ & 1 & $2.71 \times 10^{29}$ \\
Indigo   & 1.7 & $5.70 \times 10^{22}$ & 1.7 & $1.05 \times 10^{22}$ \\
Green  & 1.27 & $5.41 \times 10^{26}$ & 1.29 & $3.72 \times 10^{26}$ \\
\hline
\end{tabular}
\end{center}
\end{table}

In this section we calculate the collective contribution of normal star-forming galaxies to the EGRB that results from Eqs.~(\ref{main}) and (\ref{alltogether}). We discuss the effect on this calculation of different scaling slopes $\omega+1$ and explicit dependences on redshift entering Eq.(\ref{alltogether}) and we constrain these effects by comparing our results with Fermi EGRB data from Ackermann et al. (2015). 

For the photon spectrum $dN/dE$ we have used a broken power law with the break, in the source rest frame, at $\sim 0.6$ GeV:
\begin{equation}
\frac{dN}{dE} = \left\{ 
\begin{array}{ll}
\left(\frac{E}{0.6 {\rm \, GeV}}\right)^{-1.9},  & E< 0.6 {\rm \, GeV} \\
 \left(\frac{E}{0.6 {\rm \, GeV}}\right)^{-2.3},  & E\geq 0.6 {\rm \, GeV} 
\end{array}
\right. .
\end{equation}
For energies below $0.3 \mbox{GeV}$ we use a spectral index of $s = 1.9$ (Ackermann et al. 2012a). The slope of the high-energy branch  is consistent with the average of the spectral indices reported in the 3FGL catalog or subsequent discovery papers (see references in Table \ref{table:lums}) of detected star-forming galaxies (whether starburst galaxies are included in the sample, $s=2.33$, or not, $s=2.27$). The slope of the low-energy branch is consistent with observations of diffuse emission from the Milky Way (Abdo et al.~2009). The break at $0.6$ GeV is meant to roughly encode the pionic origin of the signal (see, e.g., Prodanovi{\'c} \& Fields 2004). Note that a much more accurate treatment of the spectrum is possible and ultimately desirable. However for the purposes of these paper (which focuses on measuring the impact of input uncertainties on the level of of the star-forming galaxy gamma-ray background contribution, rather than on making a robust prediction of its spectral shape), using a more sophisticated treatment than our simple recipe here would not affect our conclusions in any way.

The existence and location of the break at $\sim 0.3$ GeV while it is hard-coded to be at $0.6$ GeV in the source rest-frame indicates that most photons come from redshifts around $z\sim1$. 
\subsection{Effect of scaling slope $\omega+1$}\label{normals}
\begin{figure}
\includegraphics[width=1.0\columnwidth, clip]{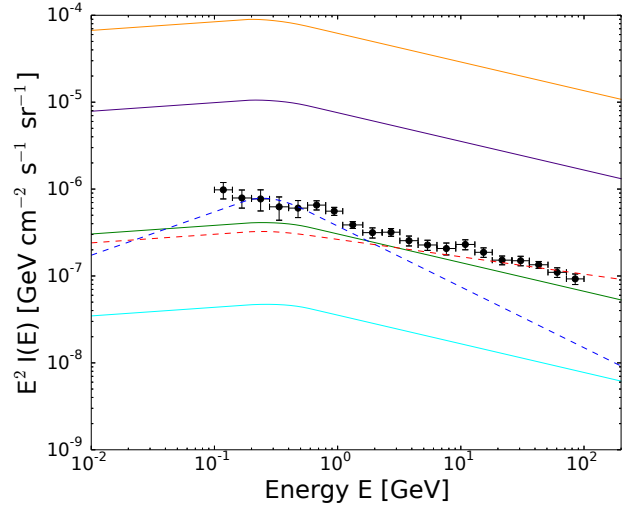}
\caption{Contribution of normal star-forming galaxies to the EGRB. \textit{Orange} line: upper bound of contribution ($\omega + 1 = 2$); \textit{solid cyan}: lower bound ($\omega + 1 =1$); \textit{indigo} line: Fields et al. (2010) theoretical prediction $\omega+1 = 1.7$.  \textit{Green} line: $\omega+1 = 1.27$, obtained using a scaling relation between $L_ {\gamma}$ and $\nu L_{\mbox{\scriptsize{NUV}}} + 2.26L_{25 \mu m}$; \textit{red dashed} line: Ackermann et al. (2012b) model; \textit{blue dashed} line: Ando \& Pavlidou (2009) model (see text for details). Fermi data are from Ackermann et al. (2015). \label{normalgalaxies}}
\end{figure}
In Fig.~\ref{normalgalaxies} we show the estimated normal star-forming galaxies contribution to the EGRB, for different values of $\omega+1$. The normalisation constant for each case is obtained using only the four normal star-forming galaxies. 

The orange line corresponds to $\omega + 1= 2$ ($x = 1$, i.e., linear correlation between the surface density of SFR and the surface density of gas). The solid cyan line corresponds to $\omega + 1= 1$ ($x = \infty$, i.e., the surface density of SFR and the surface density of gas are completely uncorrelated, or, equivalently, the gas mass does not enhance the gamma-ray emission of a galaxy). These two extreme scenarios set the bounds to the possible contribution of normal star-forming galaxies to the EGRB if all other assumptions in our calculation hold.

The green line in Fig.~\ref{normalgalaxies} is obtained using $\nu L_{\mbox{\scriptsize{NUV}}} + 2.26L_{25 \mu m}$ as a star formation tracer to determine empirically the slope of the $L_\gamma - SFR$ scaling. 
From least-square fitting we obtain $\omega + 1 = 1.27$, i.e., $L_{\gamma} \propto {\left( \nu L_{\mbox{\scriptsize{NUV}}} + 2.26L_{25 \mu m} \right)}^ {1.27}$. We examine this case because the NUV $+$ MIR (Mid-Infrared) luminosity is a better estimator of recent SFR than the TIR luminosity (Kennicutt \& Evans 2012). In this case, however, we do not consider the Milky Way in our sample, since we cannot measure its NUV luminosity. Hence, to obtain the slope, we perform least-square fitting using the three other normal star-forming galaxies (SMC, LMC, M31). Then, requiring that the $L_\gamma - L_{\rm 8-1000 \mu m}$ scaling has the same slope, we determine its normalisation as described in \S \ref{norm}. 
The indigo line is based on the formalism of Fields et al. (2010), where it is assumed that $x = 1.4$ and $\omega + 1 = 1.714$. This result is different from the result of Fields et al. (2010) since we have used the Rodighiero et al. (2010) luminosity function, and we have included the additional redshift dependence $h(z)$ to ensure that the luminosity function yields a cosmic SFR history consistent with Hopkins \& Beacom (2006). 

For comparison, we overplot the results of the Ackermann et al. (2012b) calculation (with the red dashed line) and of that of Ando \& Pavlidou (2009) (with the blue dashed line), who used the SFR density as a function of redshift instead of a luminosity function. With the exception of the spectral slopes, these models are very close to our ``fiducial'' model (green line). 

The indigo and orange lines are inconsistent with the EGRB Fermi LAT data (they over-predict the observed background). This is additional evidence that the scaling $L_{\gamma} - L_{\mbox{\scriptsize{TIR}}}$ is shallower than the theoretically predicted one based on the Kennicutt-Schmidt law (i.e., that $\omega + 1 < 1.714$). However, before we can conclude that these steep slopes are excluded, we have to test the sensitivity of the normalisation of the scaling, $L_{\gamma ,0} (\omega)$, to the number of resolved galaxies used to empirically determine its value. We do so in the next section. 

We note here that the spectral shape of the unresolved emission will not be a power law for energies $\gtrsim$ a few tens of GeV. The EGRB spectrum at higher energies is modified by three effects: i) because not all sources have the same spectral index, the hardest sources will dominate at the highest energies, giving the resulting spectrum upwards curvature (Stecker \& Salamon 1996; Pavlidou et al. 2008; Pavlidou \& Venters 2008); ii) absorption by the extragalactic background light (EBL) will eventually become important, giving the spectrum downwards curvature (e.g., Salamon \& Stecker 1998; Venters, Pavlidou \& Reyes 2009 and references therein); iii) electromagnetic cascades from the highest energy photons will alter the spectrum (e.g., Venters 2010). These effects are not treated in our calculation. 
\begin{figure}
\includegraphics[width=1.0\columnwidth, clip]{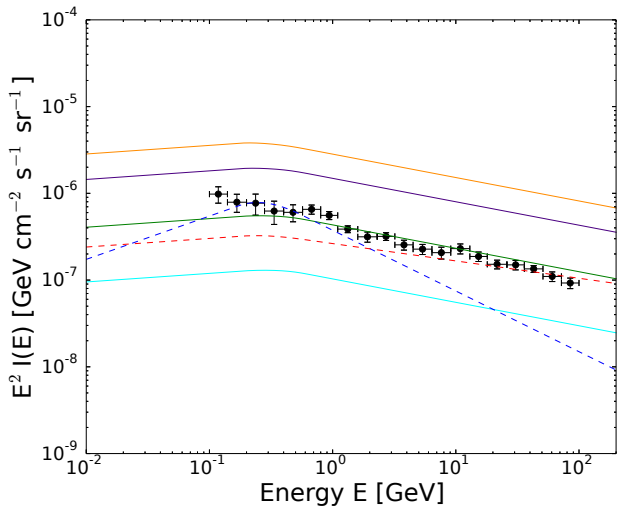}
\caption{Same as Fig.~\ref{normalgalaxies}, but now the starburst galaxies have also been considered when calculating $L_{\gamma,0}$ for each slope $\omega+1$. Fermi data are from Ackermann et al. (2015).\label{starbursts}}
\end{figure}

\subsection{Sensitivity to number of resolved galaxies}

The significant deviation (over three orders of magnitude) between predicted intensities corresponding to the extreme values of the scaling slopes $\omega+1$ in Fig.~\ref{normalgalaxies} has its origin in three factors: (a) the value of $\omega+1$ itself (i.e., the power to which an increase of the SFR at early times is raised to produce the corresponding rise in gamma-ray emissivity); (b) the suppression factor $(1+z)^{-\beta}$ due to the changing galactic disc size (higher values of $\omega$ result to lower $\beta$ and thus smaller suppression at higher redshifts); (c) the change in normalisation factor $L_{\gamma,0}$, as the same dataset is fit with power laws of different slope. 

It is expected that the effect of factor (c) above is disproportionally significant, due to the small number (four) of resolved normal galaxies that enter the normalisation factor calculation. It is therefore interesting to test whether (and to what extent) the deviation between extreme scenarios would decrease if more resolved galaxies were included in the fitting sample that is used to obtain $L_{\gamma,0}$. 

To implement this test, we repeat the calculations of \S \ref{normals}, but now we include the four starburst galaxies NGC253, M82, NGC 2146 and Arp220 in the dataset we use to determine $L_{\gamma,0}$ for each value of $\omega$.  It should be stressed that this calculation is {\em not} representative of the contribution of starburst galaxies to the EGRB, as for starburst galaxies other important assumptions in our model will not hold (for example, the scaling slope between infrared and gamma-ray luminosities should be exactly 1, and the suppression factor $(1+z)^{-\beta}$ would not enter the calculation).  

Our results are shown in Fig.~\ref{starbursts}.  Our ``fiducial'' case is only marginally affected, however the curves corresponding to the two limiting $\omega+1$ slope values now differ in predicted intensity by ``only'' 1.5 orders of magnitude. Qualitatively our results remain unchanged (the Kennicutt-Schmidt value of $\omega+1 = 1.7$ still significantly overpredicts the background), however the discrepancy is considerably smaller. We conclude that while our results clearly indicate that the level of the EGRB itself is quite constraining of the scaling slope between luminosities that trace star formation and gamma-ray luminosity, due to the extremely limited statistics of resolved, normal star-forming galaxies, we should be careful so as not to overestimate the significance of such constraints. An additional factor that could be alleviating these constraints is that at higher redshifts the transition to calorimetry may occur at lower star formation rates than today (see discussion in \S \ref{Disc}). 
\begin{figure}
\includegraphics[width=1.0\columnwidth, clip]{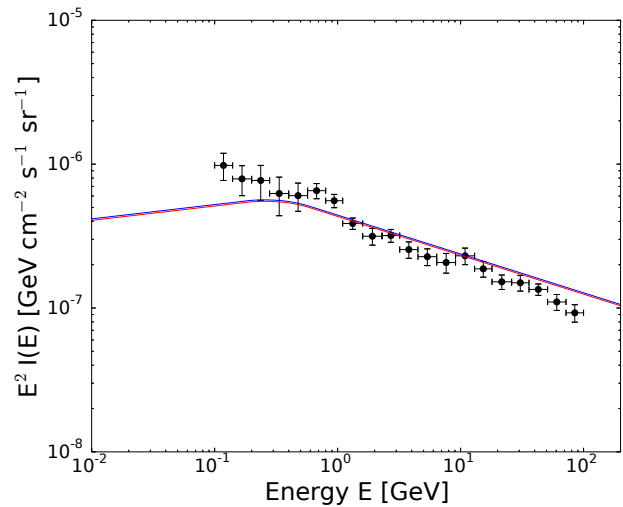}
\caption{Metallicity effect on the contribution of unresolved star-forming galaxies to the EGRB.  Blue line: ``fiducial'' model, normalisation factor $L_{\gamma,0}$ calculated using normal + starburst galaxies. Red line: same model omitting the metallicity-related factor $f(z)$. Datapoints from Ackermann et al.(2015). \label{efmetal}}
\end{figure}
\subsection{Effect of Metallicity}

The effect of including the effect of a cosmically evolving average metallicity on the calculation of the collective intensity of unresolved star-forming galaxies is shown in Fig.~\ref{efmetal}. The blue line shows our ``fiducial'' calculation with the normalisation factor $L_{\gamma,0}$ calculated including both normal and starburst galaxies. The red line shows the same calculation but omitting the metallicity-related factor $f(z)$ from Eq.~(\ref{alltogether}). The metallicity does not affect our result appreciably, thus the ratio of SNR-to-SFR can be indeed assumed to be constant for all redshifts, if the IMF is constant. It is plausible that an evolving IMF may affect this calculation mode significantly. However this case is more complicated to treat, since a redshift dependence of the IMF would have to be modeled, and then one would have to factor the changing IMF also in the scaling between infrared luminosity and SFR. 

Note that our correction is made on an average sense and does not reflect the distribution of galaxy metallicities at a given z, nor the distribution of metallicities within a single galaxy.

\subsection{Contributed intensity as a function of redshift - effect of the cosmic star formation history}
\begin{figure}
\includegraphics[width=1.0\columnwidth, clip]{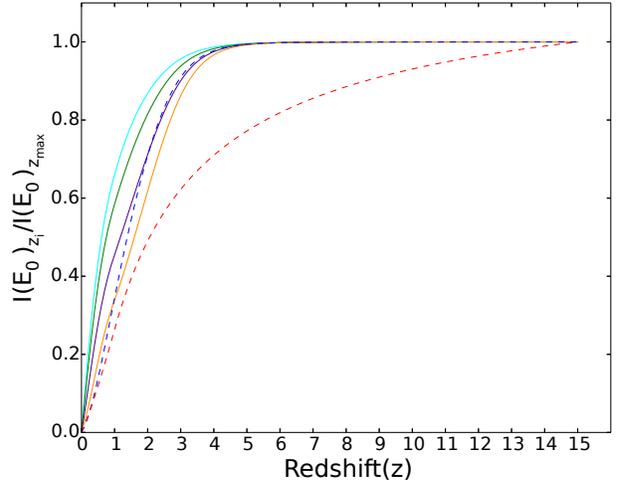}
\caption{Fraction $I(E_0)_z/I(E_0)_{z_{\rm max}}$ of intensity $I(E_0)_z$ contributed at observer-frame energy $E_0 = 0.6 {\rm \, GeV}$ by galaxies at redshifts between 0 and $z$ over total intensity $I(E_0)_{z_{\rm max}}$ contributed  by galaxies at redshifts between 0 and $z_{\rm max}$. Line color and types are as in Fig.~\ref{normalgalaxies}; note that for the red-dashed line, $z_{\rm max}=15$ in this plot and 2.5 elsewhere; see text for details.\label{redshift}}
\end{figure}

The contribution to the collective intensity from higher redshift galaxies is shown in Fig.~\ref{redshift}, where we plot the fraction $I(E_0)_z/I(E_0)_{z_{\rm max}}$ of intensity $I(E_0)_z$ contributed at observer-frame energy $E_0 = 0.6 {\rm \, GeV}$ (which corresponds to the high-energy power-law part of the spectrum) by galaxies at redshifts between 0 and $z$ over total intensity $I(E_0)_{z_{\rm max}}$ contributed  by galaxies at redshifts between 0 and $z_{\rm max}$ (the maximum redshift of our integration). Solid line colours are as in Fig.~\ref{normalgalaxies}.

For our ``fiducial'' model, over $50 \%$ of the intensity at $E_0$ comes from $z > 1$ and  $\sim 20 \%$ from $z > 2.5$. As expected, the contribution of higher redshifts is larger for higher values of $\omega$, which reflects the effect of the increase of gas mass at higher redshifts. However the overall qualitative behaviour remains the same: the cosmic star-formation history is the factor that primarily dictates the contribution of different redshifts to the total intensity. This is further emphasised by the fact that the Ando \& Pavlidou 2009 calculation (blue dashed line) also follows the general trend, although it uses very different model assumptions. The reason is that it encodes the cosmic star formation history of Hopkins \& Beacom (2006), which has been enforced in all of the models represented by solid lines through the function $h(z)$. 

As an extreme counter-example, we plot, with the red dashed line, the results for $\omega + 1 = 1.2$ (as in Ackermann et al. 2012b) but omitting the function $h(z)$ and, unlike Ackermann et al. 2012b (who take $z_{\rm max} = 2.5$), taking an extreme value of $z_{\rm max} = 15$. Because the luminosity function of Rodighiero et al. (2010) does not decline above $z=1$ and because we do not have information of the behaviour of this luminosity function at very high redshift, without the function $h(z)$ the result is sensitively dependent on the assumed value of $z_{\rm max}$. As a result, depending on the choice of $z_{\rm max}$, there can be significant contribution to the overall estimated intensity from unphysically high redshifts (in our extreme example, over $20\%$ of the intensity at $E_0$ comes from $z>5$). Conversely, a significant contribution from higher redshifts may be missed if $z_{\rm max}$ is taken to be too low. 

That the observed spectral break occurs at $\sim 0.3$ GeV while it is hard-coded to be at $0.6$ GeV in the source rest-frame indicates that most photons come from redshifts around $z\sim1$, consistent with the location of the peak of the Hopkins \& Beacom (2006) star formation history. 

\section{Summary and Discussion}\label{Disc}

The detection of several nearby normal star-forming and starburst galaxies by {\it Fermi} LAT (Ackermann et al. 2012b) has made for the first time possible to establish empirical scalings between the galaxies' gamma-ray luminosity and their luminosity at star-formation--tracing wavelengths. This development represents an unprecedented breakthrough in our ability to estimate the contribution to the EGRB from other galaxies where cosmic rays accelerated by star-formation products (supernova remnants) interact with these galaxies' interstellar gas and radiation fields. Such estimates are naturally becoming the standard in the field (e.g., Stecker \& Venters 2011; Ackermann 2012b; Linden 2017). It is therefore important to study and understand any biases that may enter these calculations, as well as identify ways that these biases can be eliminated or alleviated. This has been the scope of this work. 

We have tested whether the scaling slope between gamma-ray and infrared luminosity is well constrained, assuming that the underlying correlation has finite scatter, and given that we currently sample the underlying correlation with very few resolved galaxies. We found that even in the case where all (normal + starburst) currently resolved galaxies are considered, there is a significant ``cosmic variance'' effect: depending on whether the resolved galaxies happen to be drawn directly from the ``core'' or the ``edges'' of the correlation, the resulting fitted slopes show an appreciable spread (for scatter of 1 dec, the $1\sigma$ spread of slopes is $\sim 0.5$). 

Our estimate of the strength of this effect  is a lower limit: we have not considered errors of measurement in the luminosities; and we have not treated in any way the expected transition from a steeper slope at lower luminosities to a shallower slope at higher luminosities. This should occur because lower luminosities correspond to normal, escape-dominated star-forming galaxies, where an increased star formation rate results in a multiplicative enhancement of gamma-ray luminosity through both higher cosmic-ray acceleration and more targets (gas). On the other hand, higher luminosities correspond to starburst, calorimetric galaxies, where the gamma-ray luminosity scales as the cosmic ray flux (and hence the SFR) only. The ``true'' correlation is therefore theoretically expected to be a broken power law. In addition, the infrared luminosity where the break occurs may be redshift-dependent: due to the higher gas abundance of higher-redshift galaxies (e.g. Magdis et al. 2012), calorimetry may set in at lower star-formation rates. 

Fermi observations of the EGRB itself constrain $\omega+1$, as, for example, the theoretically predicted value of 1.7 (derived assuming escape-dominated losses and a correlation between star formation rate and gas mass given by the Kennicutt-Schmidt relation) results to an overprediction of the observed background. However, the following factors may affect these result. First, the scaling normalisation has to be derived using normal galaxies alone, as all the Fermi-observed starbursts are calorimetric (Wang \& Fields 2016\footnote{Note that Wang \& Fields (2016) in their analysis included NGC 4945 and NGC 1068 and found them to be calorimetric within errors, while they found Arp 220 exceeding that limit, an effect possibly attributable to AGN activity; as a result, the samples we treat differ somewhat from theirs.}) and will obey a different scaling. There are only four Fermi-detected normal galaxies, and hence the normalisation is very poorly constrained. Second, to obtain the contribution from normal galaxies alone, the luminosity functions have to only be integrated up to the luminosity where calorimetry sets in. In our calculations we have assumed that the result of Fields et al. 2010 that the contribution from calorimetric galaxies to this calculation is very small and can be neglected holds at all redshifts. However, that result assumed that calorimetry sets in at a fixed value of the star formation rate at all redshifts, which, as discussed above, is not obvious. 

One informative simplification we have implemented in our treatment is that, in examining the theoretically-motivated scaling slope $\omega+1=1.7$ we have allowed all galaxies, independently of star-formation rate, to contribute to the gamma-ray emission in a unified manner. Clearly this approximation ultimately breaks down, as the $1.7$ slope involves the implicit assumption that cosmic-ray losses in a galaxy are escape dominated. Once calorimetry sets in, this simple picture {\em will} break down and the slope of the gamma-ray -- star-formation-rate scaling {\em will} flatten. Fields et al. (2010) for example have explicitly excluded galaxies above some star-formation rate threshold to avoid this problem, and only include escape-dominated galaxies in their analysis. Whether the inclusion of such a cuttoff is critical for the final result is dependent on the specifics of the luminosity function adopted in the star-formation--tracing wavelengths. For the luminosity function used by Fields et al. (2010), the effect of the star-formation--rate cutoff on the gamma-ray background level was of the order of $20\%$. For the luminosity function used here (even with the $h(z)$ correction), the effect was much more pronounced, as is evident by the order-of-magnitude discrepancy between our result for $\omega+1=1.7$ and the Fields et al. (2010) result for the same slope. Additional reasons for this discrepancy include that the estimated Milky Way gamma-ray luminosity (which the authors used to normalized their gamma-ray--star-formation scaling) is low given its star formation rate, even in comparison with other local normal star-forming galaxies (see e.g. Fig. \ref{fig:scaling}) and that the luminosity function and resulting cosmic star formation history we have adopted is different.

 The magnitude of this discrepancy in the case we have treated here, however,  emphasizes that a continuous, unified treatment of escape-dominated and calorimetric galaxies should ideally be used in predicting their contribution to the diffuse gamma-ray emission.
  
We have found that the cosmic evolution of the supernova mass cutoff due to an increasing average metallcity does not appreciably affect this calculation, and can safely be neglected. 

On the other hand, we have found that the different and redshift-dependent performance of various star formation tracers can considerably affect the calculation. For example, although the Kennicutt (1998b) scaling between total infrared luminosity and star-formation rate holds only for optically thick galaxies, it is uniformly applied to all sources, and infrared luminosity functions have been commonly used due to their availability rather than the robustness with which they trace star formation. One possible solution to this problem, which we have applied here, is to apply a redshift-dependent correction factor to the infrared luminosity function that brings the resulting cosmic star formation history in agreement with combination estimates that use all available star formation tracers, such as the one by Hopkins \& Beacom (2006). 

An evolving stellar IMF is another possible bias that may affect the ERGB calculation, however we have not treated such an evolution in this work. Instead, we used a Salpeter IMF (Salpeter 1955) for all redshifts. A proper treatment of an evolving IMF should model its impact on the scaling between star-formation-tracing luminosity and SFR, as well as on the SNR - SFR scaling. 

We note that while our results indicate a rather robust spectral break (a peak in a $\log (E^2I)$-$\log E$ plot) tracing the redshift of the peak in cosmic star formation history, such a feature is not visible in Fermi data. This may be an indication that gamma rays from star-forming galaxies, even if they are a significant contribution to the EGRB (Linden 2017), are not the dominant one. A contribution of normal galaxies to the EGRB low enough that the spectral peak would be ``hidden'' in the overall observed spectrum is clearly well within the parameter space allowed by other observational inputs in our model. 

We conclude that empirical scalings between star-formation tracing luminosities and gamma-ray luminosities of star-forming galaxies, despite being a significant breakthrough that allows the construction of gamma-ray luminosity functions, do not constrain the latter enough to produce precision estimates of the normal-galaxy contribution to the EGRB. Once all possible biases are considered, these estimates are at best accurate within an order of magnitude, with the main limitation being that the star-forming galaxies that can be individually resolved in gamma rays are (a) very few and (b) all part of the local Universe. Further progress towards a more accurate determination of the normal galaxy contribution to the EGRB necessarily passes through more sophisticated models incorporating all recent progress in our understanding of a cosmically-evolving star-formation process, and a unified treatment of normal and starburst galaxies, including the gradual transition to calorimetry.

\section*{Acknowledgements}

We are grateful to Xilu Wang, Brian Fields, Floyd Stecker, and an anonymous referee for helpful comments that helped improve this work. I.K. would like to thank A. Tritsis for useful suggestions and discussions.  Support for this work was provided by the European Research Council under the European Unions Seventh Framework Programme (FP/2007-2013) / ERC Grant Agreement n. 617001 and by a Marie Sklodowska-Curie RISE grant "ASTROSTAT" (project number 691164).

\end{document}